# Preparation of Layered Organic-inorganic Nanocomposites of Copper by Laser Ablation in Water Solution of Surfactant SDS.

**V. T. Karpukhin, M. M. Malikov[*], T. I. Borodina, G. E. Val'yano, and O. A. Gololobova**

*Joint Institute for High Temperatures RAS, Moscow, Russia.*

E-mail: *mmalikov@oivtran.ru

The data experimental synthesis and studies of layered organic-inorganic nanocomposites [$Cu_2(OH)_3$ + DS], resulting from ablation of copper in aqueous solutions of surfactant – dodecyl sodium sulfate (SDS) are presented. By the methods of absorption spectroscopy of colloidal solutions, X-ray diffraction, scanning electron (SEM) and atomic force microscopy (AFM) of solid phase colloids was traced the formation dynamics of these composites, depending on the exposure duration of copper vapor laser radiation on the target of copper as well as the aging time of the colloid. Bilayered structures of composite [$Cu_2(OH)_3$ + DS] fabricated by method of laser ablation copper metal target in liquid are demonstrated for the first time.

## 1. Introduction

Layered organic-inorganic hybrid nanocomposites are of significant interest to researchers in terms of their applications in science and technology. These include a broad class of chemical compounds such as (1) – a layered double hydroxides (LDH), (2) – hydroxy double salts (HDS's) and (3) – hydroxides of metals, in which various organic anions were intercalated in the space between the layers. The structural variability of these materials leads to the appearance of new chemical and physical properties – variable magnetism [1, 2], efficient catalysis and high ion-exchange capacity [3]. The use of such nanocomposites promises to improve the mechanical and thermal stability of polymers in which composites have been dispersed [4], and also opens up the possibility for a creation of new optoelectronic devices (stochastic lasers, LEDs [5, 6], sensors [7], etc).

Synthesis and analysis of chemical properties of layered nanocomposites devoted hundreds of publications [3]. In the last decade for the synthesis of metal, oxide, hydroxide nanostructures used method of laser ablation of metals in a liquid medium [8]. However, the researches aimed at producing layered organic-inorganic composites by this method are not enough.

In this paper the authors present the results of studying the synthesis layered nanocomposites of transition metal copper. The synthesis was carried out by laser ablation of metal target in aqueous solutions of surfactant. The synthesized material belongs to the second group composites. Its structural formula is as follows: $(M)_2(OH)_3 X * z * H_2O$, where M – divalent metals (Cu) and X – intercalated anion – alkyl sulfate ($C_nH_{2n+1}SO_4^-$), where n = 12.

In this paper as a source of radiation was chosen copper vapor laser (CVL) with an output power $P_{out}$ – 10–15 W, pulse duration $\tau_p$ = 20 ns and pulse repetition frequency $f_p$ = 10 kHz.

Time of irradiation of metal target and colloid in this case is up to 1000 times higher than the same time in the most of the experiments carried out with Nd:YAG lasers ($P_{out}$ = 1 W, $\tau_p$ = 5–10 ns, $f_p$ = 10 Hz). The specificity of using CVL determines the nature of the synthesis of nanostructures and quality of finite products.

## 2. Description of the Experiment

The technique for the laser ablation of materials in liquid media was described in details in a number of original papers and overviews [9, 10]. In this experiment, the generation of CVL was performed at two wavelengths – 510 and 578 nm; the ratio of radiation power in the lines was, respectively, 2:1. The laser beam was focused on the surface of a target by achromatic lens with a focal length f = 280 mm, which ensured a spot size of less than 100 microns. The target was placed in a cell with deionized water or aqueous solutions of surfactant. The volume of liquid in the cell was ~ 2 $cm^3$. The cell was placed in the vessel with the cooling water which temperature is maintained at 330 K. The vessel was mounted on a movable stage, allowing continuously move the focal spot on the target surface. Surfactant used in the experiment – SDS ($C_{12}H_{25}SO_4Na$) is an anionic surfactant.

Optical characteristics of the obtained colloidal solution containing nanostructure of Cu, $Cu_2O$, CuO and other compounds were analyzed by the method of optical absorption, with spectral range from 200 to 700 nm, on a spectrophotometer SF-46 with automatic data processing system. The structure and composition of the solid phase colloidal solution prepared by centrifugation at 4000 rpm and dried at a temperature of 320–330 K were investigated by X-ray diffractometer DRON-2 (Kα line of copper). The shape and size of nanostructures were studied with an atomic force microscope (AFM) Solver P47-PRO (in semicontact topography mode) and scanning electron microscopy (SEM, Hitachi S405A, 15kV). The samples for AFM analysis were prepared by single or multiple coats of wet precipitate obtained after centrifugation on a glass slide with a follow-drying the precipitate at a temperature of 312–323 K under atmospheric conditions. The number of layers of precipitate was adjusted empirically to achieve a sufficient sharpness of X-ray diffraction patterns and AFM images.

## 3. The Experimental Results and Discussion

Figure 1 shows the absorption spectra of copper colloids. In this spectra there are the sharp and wide peaks (curves *1, 2, 3*), reflecting the presence in colloids of the copper oxides nanostructures – CuO and $Cu_2O$. At numerous studies of synthesis of copper nanoparticles by chemical and other methods, were found similar features of the spectra and were discussed theirs nature. In [11] the increase the absorption of 260 nm and 340 nm, was associated with interband transitions in nanocrystallites $Cu_2O$. In these zones are located and interband transitions of copper ions [12]. A weak peak around 630–640 nm was interpreted as the exciton absorption of $Cu_2O$ (band gap ≈ 2eV), the peak at λ ≈ 580 nm was recorded in [13]. It was assumed that the increase in absorption can occur due to plasmon resonance nanoparticles of copper and copper ion interband transitions [14]. As in the experiments with Zn [10], the spectra of colloids containing different concentrations of SDS (curves *4, 5, 6, 7*) considerably distinguish from water colloids. The overall level of absorption increases, especially in the long wavelength (> 450 nm) spectral region. But the spectra of colloid with 0.01 M SDS is quite different from others. The degree of decreasing of absorbance at wavelengths of 350 nm is much smaller than that of other colloids and it is characterized by the presence of large-scale fractal structure in this colloid. This fact is confirmed by a significant decreasing of absorption at the aging and after the centrifugation of colloid.

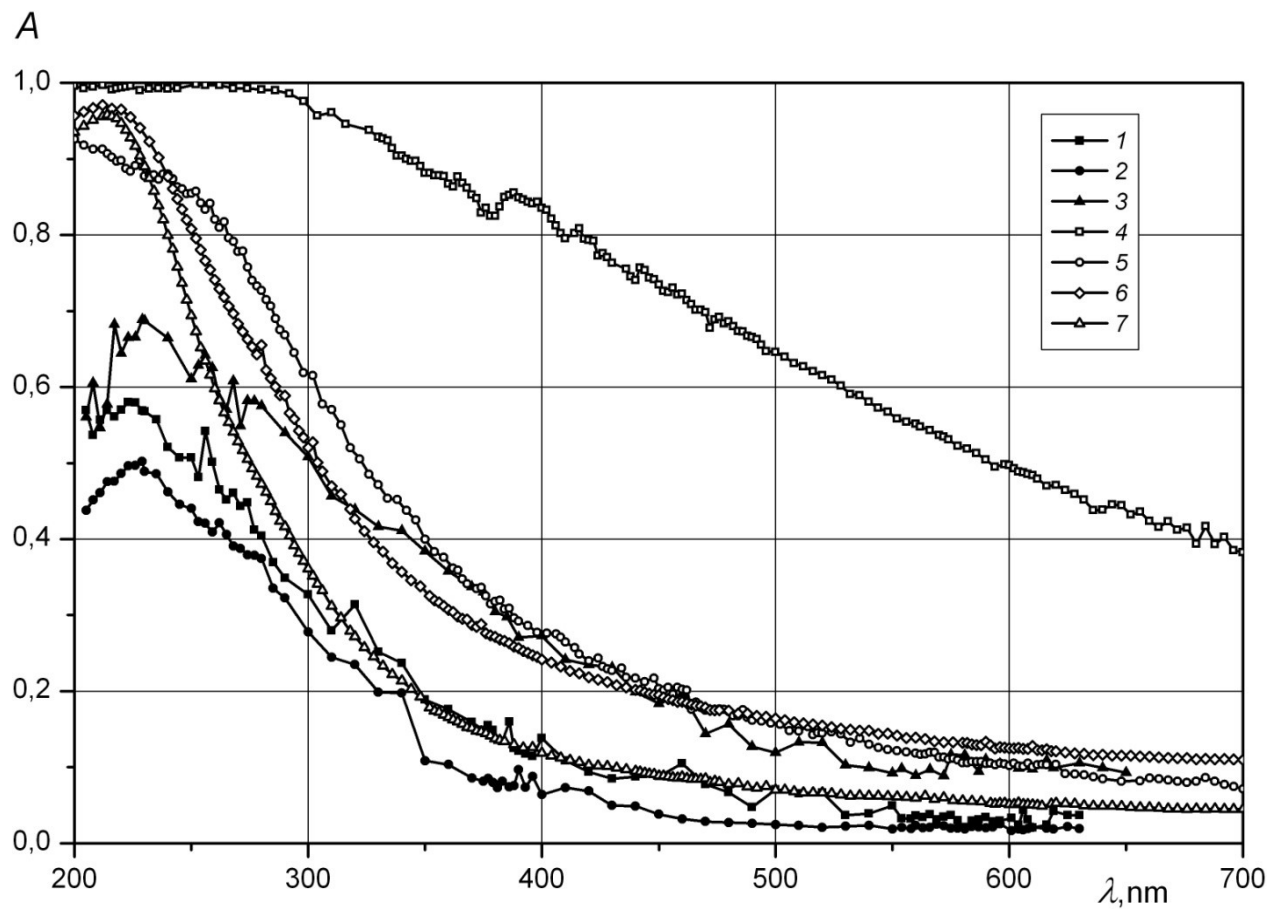

Figure 1. The absorption spectra $A(\lambda)$: 1–3 – Cu + $H_2O$; 1 – $\tau_{ag}$ = 2 min, $\tau_e$ = 40 min; 2 – $\tau_{ag}$ = 3 h, $\tau_e$ = 40 min; 3 – $\tau_{ag}$ = 1.3 h, $\tau_e$ = 3 h ; 4–5 – Cu + 0.01 SDS; 4 – $\tau_{ag}$ = 10 min, $\tau_e$ = 3 h; 5 – $\tau_{ag}$ = 64 h, $\tau_e$ = 3 h; 6–7 – Cu + 0.1 SDS; 6 – $\tau_{ag}$ = 10 min, $\tau_e$ = 3 h; 7 – $\tau_{ag}$ = 19 h, $\tau_e$ = 3 h.

XRD patterns, AFM and SEM images shows that solid phase of 0.01M SDS colloids consist of a two-dimensional layered structures with two alkyl sulfate anion layers accommodated between the two cupric hydroxide layers. The formula of this complex is $[Cu_2(OH)_3DS^- – Cu_2(OH)_3CH_3(CH_2)_{11}OSO_3]$. XRD patterns (fig. 2) demonstrate the two layered structures with the basal spacing $d$ = 3.887 nm and $d$ = 3.27 nm. Dimensions of the interlayer space in this experiment shows that the resulting structures are bilayered. Indeed, as follows from the estimates given in [15], at interlayer size of 3.887 nm can with reasonable certainty meet a bilayered structure with layers of DS anions arranged either in a partially interdigitated belayred or tilted to the layers of copper hydroxide at an angle about 71°. The same structure with $d \approx 3.92$ nm was obtained by chemical way in [4]. Size of 3.27 nm may be a structure in which the alkyl chains are oriented at an angle about 48° to the hydroxide layers or a partially interdigitated.

From the analysis of SEM images (fig. 3) and AFM images (fig. 4) can conclude that the observed flat structures (platelets) are the «sandwiches», folded from many bilayers of composite. It should be noted that the above structure of organic-inorganic composite of copper, apparently, by laser ablation in liquids was obtained for the first time.

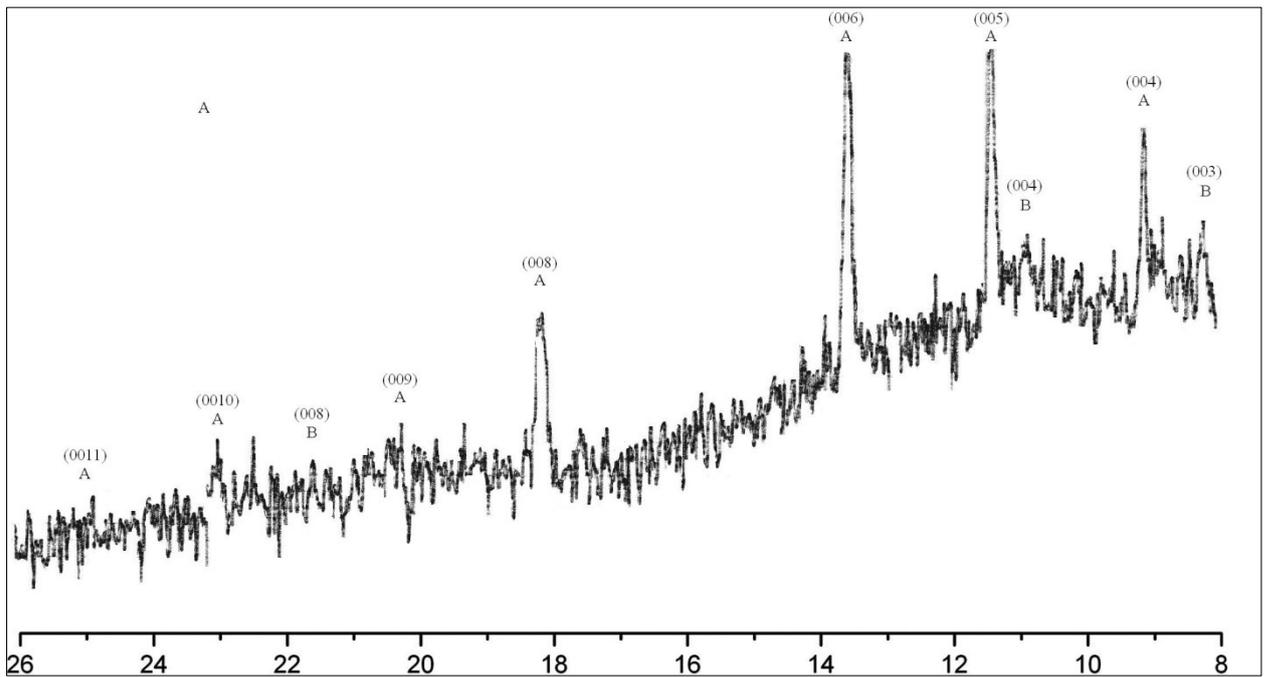

Figure 2. X-ray patterns of sediment extracted from the colloidal solution Cu + 0.01 M SDS. Phases: A – the basal spacing $d = 3.887$ nm, B – $d = 3.27$ nm.

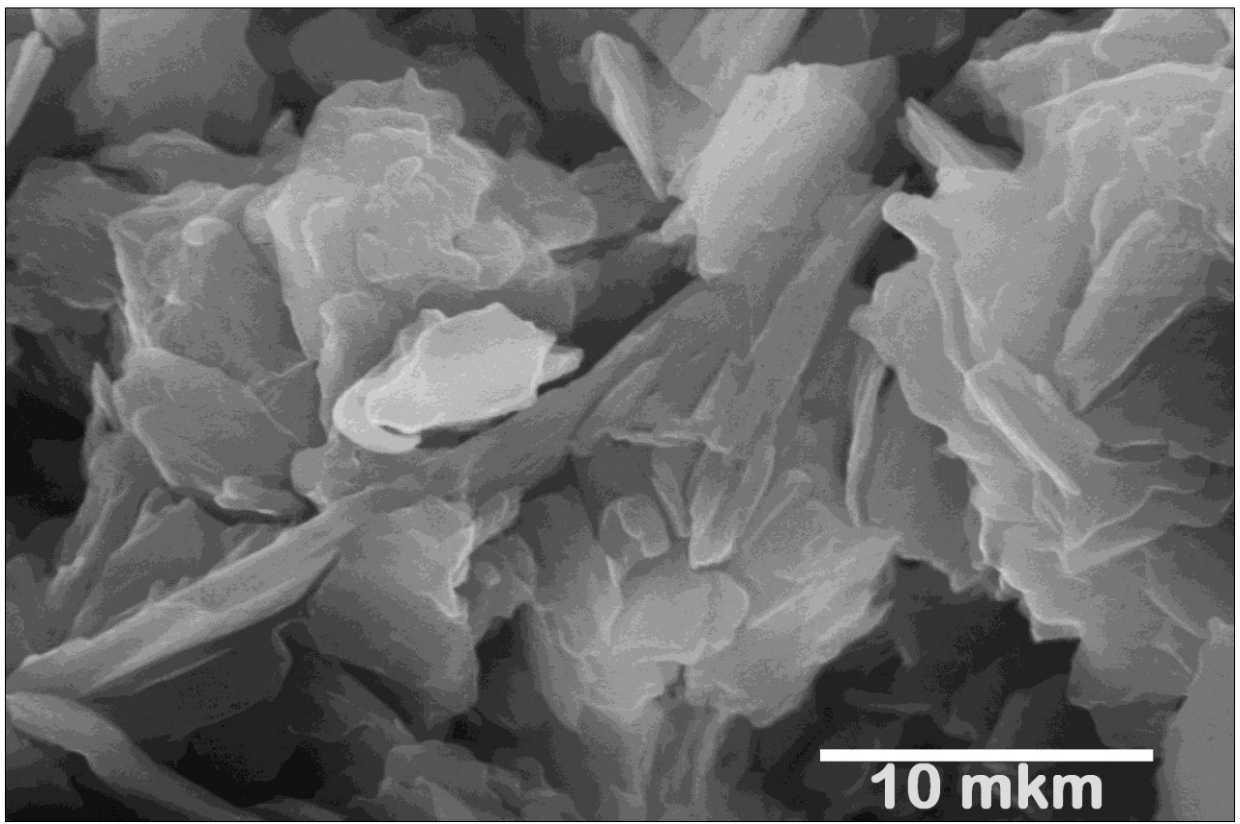

Figure 3. SEM image of layered structures of the preparation obtained after laser ablation of copper target in the 0.01M solution of SDS.

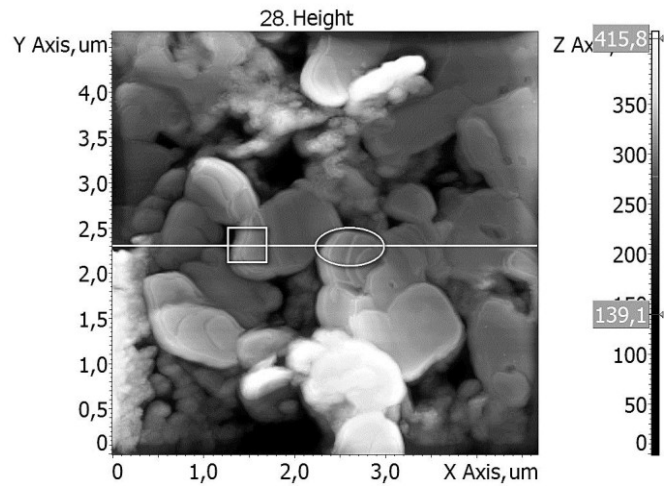

(a)

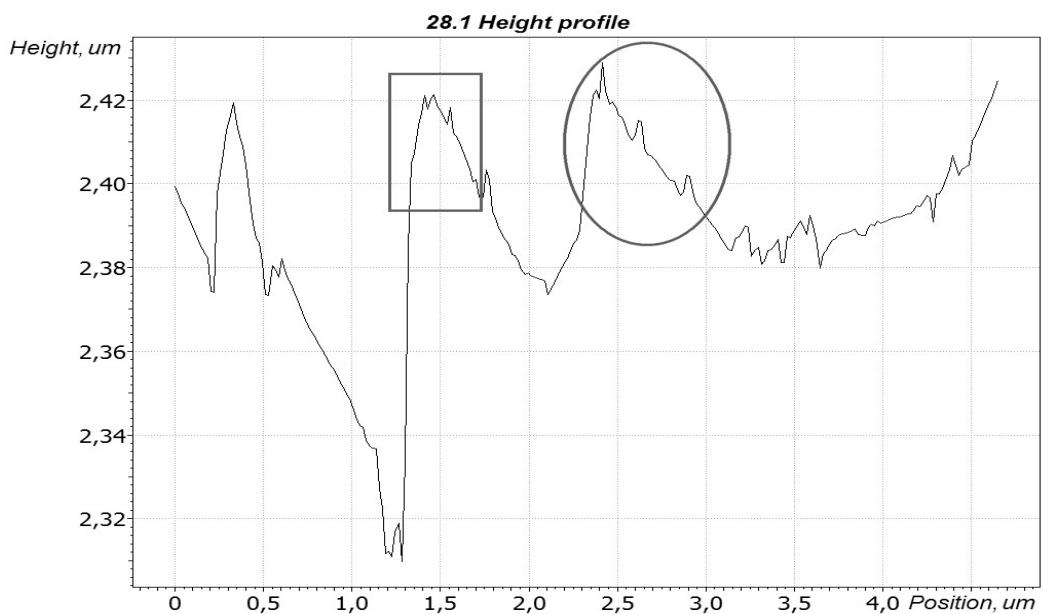

(b)

Figure 4. AFM image of part of the preparation obtained after laser ablation of copper target in the 0.01M solution of SDS, (a) – the image of structure, (b) – the trace of cantilever AFM (the profile of «sandwich» edges).

## 4. Conclusion

The optical spectra, diffraction patterns and AFM images reflect the dynamics of formation of nanostructures due to the use in the experiment of a powerful copper vapor laser. The increase in the total ablation time and time colloid irradiation at high pulse repetition rate, result in a rise of colloid temperature due to the high average power of irradiation and lead to an intense synthesis of nanoparticles on the basis of copper (oxides, hydroxides, layered organic-inorganic composites), the emergence of clusters and large (up to hundreds of nanometers or more) complexes with different structures and forms – fractal aggregates. Based on the theory of fractals is possible qualitative explanation of the features of optical spectra of the investigated colloids [10, 16]. Specificity of the use of surfactants for the ablation process is demonstrated the formation of micelle, consisting of surfactant molecules and nanoparticles. On the one hand, the surfactant molecules, surrounding the nanoparticles, are limited their growth and aggregation, on the other – there is a possibility of synthesis of various chemical compounds on the basis of the surfactant, water and metal. In these experiments two different shapes of bilayered copper composites [$Cu_2(OH)_3DS$] was prepared and these composites, probably, are obtained by laser ablation in water solution of SDS for the first time.